\documentstyle[equations,12pt]{article}
\begin{document}

\begin{flushright}
hep-th/9512172 \\
IPNO/TH 95--73
\end{flushright}
\vspace{1 cm}
\begin{center}
{\large {\bf There are no abnormal solutions of the Bethe$-$Salpeter
equation in the static model}}
\vspace{1 cm}

Hassen JALLOULI\footnote{e-mail: jallouli@ipncls.in2p3.fr} and
Hgop SAZDJIAN\footnote{e-mail: sazdjian@ipncls.in2p3.fr}
 \\

\renewcommand{\thefootnote}{\fnsymbol{footnote}}
{\it Division de Physique Th\'eorique\footnote{Unit\'e de Recherche des
Universit\'es Paris 11 et Paris 6 associ\'ee au CNRS.},
Institut de Physique Nucl\'eaire, \\
Universit\'e Paris XI, \\
F-91406 Orsay Cedex, France}
\vspace{3 cm}

{\large Abstract}
\end{center}
The four-point Green's function of static QED, where a fermion and an
antifermion are located at fixed space positions, is calculated in
covariant gauges. The bound state spectrum does not display any abnormal
state corresponding to excitations of the relative time. The equation
that was established by Mugibayashi in this model and which
has abnormal solutions does not coincide with the Bethe$-$Salpeter
equation. Gauge transformation from the Coulomb gauge also confirms
the absence of abnormal solutions in the Bethe$-$Salpeter equation.

\newpage

One of the puzzling features of the Bethe$-$Salpeter equation
\cite{sb,gml,n1} is the appearance in the bound state spectrum of abnormal
solutions corresponding to excitations of the relative energy (or
relative time) \cite{n1,n2}. These states do not have nonrelativistic
counterparts, neither do they appear in the hamiltonian formalism.
The first example of such states was provided by the Wick$-$Cutkosky
model \cite{w,c}, where, in the one scalar photon exchange
approximation, an infinite number of new states, with a new quantum
number, appeared  in the bound state spectrum for values of the coupling
constant greater than a critical value. It was also established that
these states did couple to the two-particle Green's function \cite{n3}.
\par
The fact that the presence of abnormal states is due to the excitations
of the relative time was checked by the analysis of the problem in the
static model \cite{otu}, in which two fermions are held fixed at definite
positions in space. In this case the whole set of excited states should come
entirely from the relative time excitations. It was again found that in the
one-photon exchange approximation an infinite number of abnormal states
exist when the coupling constant exceeds a critical value.
\par
It was suspected \cite{w} that the appearance of abnormal solutions might
be due to the one-photon exchange approximation and that multiphoton
exchange contributions might remove them from the spectrum. In this
connection the exact static model was analyzed by Mugibayashi \cite{m}. Using
hamiltonian formalism he derived a second order differential equation
that he identified with the Bethe$-$Salpeter equation. This equation
had again an  infinite number of abnormal states (for massless photons);
the situation was even worse than before, since these were now
appearing for {\it any} value of the coupling constant.
\par
The trouble with Mugibayashi's result is that it is in obvious
contradiction with experimental data: no such states are observed in
QED. [Mugibayashi's scalar interaction can be transposed to QED.] On the
other hand, the one-photon approximation results might still escape
experimental tests, since the QED coupling constant $\alpha$ is far
below from the critical value at which abnormal states appear.
\par
These troublesome results and conclusions have led us to reanalyse
the static model. The key ingredient in our analysis is the fact that
the Green's function of the static model is exactly calculable and
hence an explicit check of its poles provides indication about the
bound state spectrum. Only one pole, that of the (normal) ground state,
is found. The contradiction that exists with Mugibayashi's result stems
from the fact that his equation is {\it not} the Bethe$-$Salpeter
equation, but rather a secondary equation of the theory. Calculating
the Bethe$-$Salpeter kernel from Feynman diagrams or from the inverse
of the Green's function, it is found, contrary to Mugibayashi's
observation, that the kernel contains, in covariant gauges, an infinite
series of multiphoton exchange diagrams.
\par
The above results are also verified in the Coulomb gauge, where the
Bethe-Salpeter equation has only one bound state. Gauge invariance of
the theory can then be used to provide additional justification to
the absence of abnormal solutions of the Bethe-Salpeter equation in
other gauges.
\par
We therefore conclude that the Bethe$-$Salpeter equation does not have
any abnormal solution in the exact static model. Since it is unlikely
that excitations of the space coordinates induce by themeselves relative
time excitations, one should expect that this result remains also true in
the more general four-dimensional theory. Our result confirms Wick's
conjecture \cite{w} that abnormal states are spurious solutions due
to the one-photon exchange approximation.
\par
The absence of abnormal solutions in the static model has also its
relevance for Heavy Quark Effective Theory, which is formulated as
an expansion around the static model \cite{fclght}.
\par
We now turn to the details of the calculations. We consider spinor
electrodynamics in the static limit where one fermion and one
antifermion are held fixed at definite space positions.
The corresponding lagrangian density, in an arbitrary linear covariant
gauge characterized by a parameter $\xi$, is:
\begin{eqnarray} \label{e1}
{\cal L}&=&\overline\psi _1  (i\gamma_0 \partial ^0 - m_1 - g_1 \gamma_0 A^0 )
\psi _1 +
\overline\psi _2  (i\gamma_0 \partial ^0 - m_2 - g_2 \gamma_0 A^0 ) \psi _2
\nonumber \\
& &\ \ \ \ \ \ \ \ \ \ \ -
\frac{1}{4} F_{\mu\nu}F^{\mu\nu}-\frac{1}{2(1-\xi)}(\partial_\mu A^{\mu})^2
\ .
\end{eqnarray}
Since the spacelike components of the $\gamma$-matrices are absent, the
fermion fields can be classified according to the eigenvalues of the
matrices $\gamma_0$ ($+1$ for the fermion, $-1$ for the antifermion).
In the Feynman gauge, the above lagrangian density becomes equivalent,
for its mutual interaction part,
to that of the scalar interaction considered in Ref. \cite{m}.
\par
The four-point Green's function describing the scattering of the two
particles is defined, up to a normalization factor, by the functional
integral:
\newpage
\begin{eqnarray}  \label{e2}
G(x_1,x_2,x_1',x_2')&=& \int D\psi D\overline\psi DA
\psi_1(x_1) \psi_2(x_2')
\overline \psi_2(x_2) \overline\psi_1(x_1')
e^{iS(\psi,\overline\psi,A)+i\int J_0A^0} \nonumber  \\
&\equiv&<\psi_1(x_1) \psi_2(x_2')
\overline\psi_2(x_2) \overline\psi_1(x_1')
e^{iS(\psi,\overline\psi,A)+i\int J_0A^0} > \ ,
\end{eqnarray}
where $S=\int d^4 x {\cal L}(x)$ is the action and
$J_0$ is an external source for the field $A^0$. To calculate this
integral we eliminate the interaction term in $S$ by the
following transformations of the fermion fields:
\begin{equation}  \label{e2p}
\psi_i(x)\rightarrow e^{-ig_i \frac {1}{\partial ^0}A^0(x)}\psi_i(x)\ ,
\ \ \ \ \ \ \
\overline\psi_i(x)\rightarrow e^{ ig_i \frac {1}{\partial ^0}A^0(x)}
\overline\psi_i(x)\ \ \ \ \ (i=1,2)\ ,
\end{equation}
where $\frac{1}{\partial^0} A^0(x)$ designates a primitive of $A^0(x)$
with respect to $x_0$, $\frac {1}{\partial ^0}$ being defined in momentum
space as the principal value of $i/k^0$.
The Jacobian of these transformations is unity and we obtain for $G$:
\begin{eqnarray} \label{e2s}
G&=&<\psi_1(x_1)e^{-ig_1 \frac{1}{\partial ^0}A^0(x_1)}
\psi_2(x_2')e^{-ig_2 \frac{1}{\partial ^0}A^0(x_2')}
\overline\psi_2(x_2)e^{ig_2 \frac{1}{\partial ^0}A^0(x_2)}
\overline\psi_1(x_1')e^{ig_1\frac {1}{\partial ^0}A^0(x_1')}\nonumber \\
& &\ \ \ \ \ \ \ \ \ \ \ \ \ \ \ \times e^{iS_0(\psi,\overline \psi,A)+
i\int J_0A^0}> \ ,
\end{eqnarray}
where $S_0$ is the free action .
\par
The remaining integration in the functional integral is gaussian and yields
in a straightforward way:
\begin{equation}  \label{e3}
G\ =\ G_0
e^{\frac{i}{2} \int \frac {d^4k}{(2\pi)^4}\ \frac{1}{k^2+i\epsilon}
\ (g_{00}-\xi \frac{k_0^2}{k^2+i\epsilon})\ \left| \widetilde J^0(k)+J^0(k)
\right |^2}\ ,
\end{equation}
where $G_0$ is the free two-particle propagator,
\begin{eqnarray} \label{e3p}
G_0 &=& G_{10} G_{20}\ ,\nonumber \\
G_{i0}(x_i-x_i')&=& \theta(x_i^0-x_i^{\prime 0})
e^{-im_i(x_i^0-x_i^{\prime 0})}\delta^3({\bf x}_i-{\bf x}_i')\ \ \ \ \ \ \
(i=1,2)\ ,
\end{eqnarray}
and $\widetilde J^0$ is:
\begin{equation} \label{e4}
\widetilde J^0(k,x_1,x_2,x_1',x_2')\ =\ i\frac{1}{k_0}
\big[g_1( e^{ik.x_1}-e^{ik.x_1'}) - g_2(e^{ik.x_2}- e^{ik.x_2'})\big]\ .
\end{equation}
For $J^0=0$ one has:
\begin{equation} \label{e4s}
G\ =\ G_0e^{\frac{i}{2}\int \frac{d^4k}{(2\pi)^4}
(1-\xi\frac{k_0^2}{k^2+i\epsilon})\frac{\left | g_1(e^{ik.x_1}-e^{ik.x_1'})
-g_2(e^{ik.x_2}-e^{ik.x_2'})\right | ^2}{(k^2+i\epsilon) k_0^2}}\ .
\end{equation}
In the argument of this exponential the terms proportional to $g_i ^2\
(i=1,2)$ contribute to the mass and wave function renormalizations.
After carrying out the latter renormalizations (we shall continue
denoting by the same notations the renormalized masses) and setting
\begin{equation} \label{e4p}
g_1 = g_2 = g\ ,\ \ \ \ \ \alpha\ =\ \frac{g^2}{4\pi}\ ,\ \ \ \ \
\lambda \ =\ \frac{g^2}{4\pi^2}\ ,
\end{equation}
we obtain from Eq. (\ref{e4s}) for the renormalized Green's function:
\begin{eqnarray} \label{e5}
G &=& G_0 e^{ i\big[ \chi_{\xi}(x_1-x_2) - \chi_{\xi}
(x_1-x_2') - \chi_{\xi}(x_2-x_1') + \chi_{\xi}(x_1'-x_2')\big]}
\nonumber \\
& & \ \ \ \ \ \ \ \ \ \times e^{i\big[h_{1\xi}(x_1^0-x_1^
{\prime 0}) + h_{2\xi}(x_2^0-x_2^{\prime 0})\big]}\ ,
\end{eqnarray}
where $\chi_{\xi}$ and $h_{i\xi}$  $(i=1,2)$ are defined as:
\begin{eqnarray} \label{e5p}
\chi_{\xi}(x) &=& \lambda \int \frac{d^4k}{(2\pi)^4}\ \frac
{e^{i{\bf k.x}}}{k^2+i\epsilon}\ (1-\xi\frac{k_0^2}{k^2+i\epsilon})\
\frac{(e^{-ik_0x^0}-1)}{(-k_0+i\epsilon)(k_0+i\epsilon)}\nonumber \\
&=& -i\lambda\ \big\{\ \frac{x^0}{2r}\ln\left(\frac{r+x^0+i\epsilon}
{r-x^0+i\epsilon}\right) + \frac{(\xi+2)}{4}\ln\left(\frac{-x^2+i\epsilon}
{r^2}\right)\big\}\ ,\ \ \ \ \ r=|{\bf x}|\ ,\nonumber \\
& &\\
h_{i\xi}(x^0) &=& -i\lambda \frac{(\xi+2)}{4}\ln(-m_i^2 (x^{02}-r_0^2)
+i\epsilon)\ \ \ \ \ \ (i=1,2)\ ,
\end{eqnarray}
where $r_0$ ($\simeq 0$) is the renormalization short distance regulator.
Notice that $\chi$ is an even function of $x^0$. The functions $h_{i\xi}$
come from the one-fermion propagator renormalizations. They represent
the cloud of scalar and longitudinal photons usually appearing in
covariant gauges \cite{jz}. Their effect in momentum space consists in
replacing the pole of the propagator by a power law singularity.
A similar effect is also contained in the second term of $\chi_\xi$.
\par
In the absence of radiative corrections ($h_{\xi}=0$),
formula (\ref{e5}) can also be obtained by directly summing the Feynman
diagrams corresponding to the multiphoton excahanges and contributing to the
fermion-antifermion scattering amplitude. Because of the linearity of
the denominators of the fermion propagators in momentum space [Eqs.
(\ref{e3p})], the summation technique of the eikonal approximation
\cite{cwls} can be applied in exact fashion, with the fermions kept
off their mass shell. When radiative corrections are present, the
calculations are more involved. The emission amplitude of $n$ photons
from a fermion line is completely determined, through successive uses
of Ward-Takahashi identities, in terms of the fermion propagator. After
this step, the eikonal summation technique can be applied with
appropriate adaptations and formula (\ref{e5}) is reached.
\par
In the following we shall use for the total and relative variables the
notations:
\begin{eqnarray} \label{e6}
X &=& \frac{x_1+x_2}{2}\ ,\ \ \ \ X'\ =\ \frac{x_1'+x_2'}{2}\ ,\ \ \ \
t\ =\ X^0-X^{\prime 0}\ ,\nonumber \\
x &=& x_1-x_2\ ,\ \ \ \ x'\ =\ x_1'-x_2'\ , \ \ \ \ r\ =\ |{\bf x}_1-
{\bf x}_2|\ =\ |{\bf x}_1'-{\bf x}_2'|\ ,\nonumber \\
P &=& p_1+p_2\ ,\ \ \ \ p\ =\ \frac{1}{2}(p_1-p_2)\ .
\end{eqnarray}
\par
To extract from the Green's function (\ref{e5}) the bound state spectrum
of the theory we let the time $t$ tend to infinity, keeping $x^0$ and
$x^{\prime 0}$ fixed. Because of the
cluster decomposition \cite{gml}, the Green's function behaves in general
for infinite time separation as:
\begin{equation}  \label{e7}
G(x_1,x_2,x_1',x_2')_{\ \buildrel {\displaystyle \sim }\over {t \to
\infty}} \ \sum _n \phi_n(x_1,x_2) \overline \phi_n(x_1',x_2')\
=\ \sum _n \phi_n(x)e^{-iP_{0n}t}\ \overline \phi_n(x')\ .
\end{equation}
In this limit the Green's function (\ref{e5}) yields only one
exponential factor and behaves as:
\begin{equation}   \label{e8}
G_{\ \buildrel {{\displaystyle \sim }}\over {t \to \infty}} \
e^{-i(m_1+m_2-\alpha/r)t}\ .
\end{equation}
This behavior signals the presence of one bound state with energy
\begin{equation} \label{e8p}
P_0\ =\ m_1+m_2-\frac{\alpha}{r}\ ,
\end{equation}
which is gauge invariant. Notice that the contributions of the scalar
and longitudinal photon clouds in the functions $\chi_\xi$ and $h_{i\xi}$
have cancelled each other in the limit $t\rightarrow \infty$. This is
a consequence of the electric neutrality of the bound state considered
here.
\par
The above result about the bound state spectrum can also be obtained by
analyzing the Green's function in total momentum space. After the
isolation of the ground state pole (\ref{e8p}), no other infinite type
singularities appear.
\par
We conclude from this calculation that the static model has only one bound
state, the normal one, and no abnormal states exist.
\par
We next turn to the analysis of the problem  by means of the Bethe-Salpeter
equation. By functional methods \cite{s} one establishes the following
equation for the Green's function:
\begin{equation}  \label{e10}
(i\gamma_{10} \partial_{x_{10}} -m_1)G(x_1,x_2,x_1',x_2')\ =\
i\delta^4(x_1-x_1') G_{20}(x_2-x_2') -
ig\gamma_{10}\frac{\delta}{\delta J_0(x_1)}G(x_1,x_2,x_1',x_2')\ ,
\end{equation}
and a similar one with the interchanges $1\leftrightarrow 2$. Here
$G_{i0}$ ($i=1,2$) is the free propagator of fermion $i$ [Eq.(\ref{e3p})].
The complete expression of the Green's function [Eq. (\ref{e3})] can
then be used to transform these equations into integro-differential
equations for $G$. Equivalently, one can directly apply the operators
$G_{i0}^{-1}$ ($i=1,2$) on the expression (\ref{e5}) of $G$. One
obtains the following two differential equations (we omit radiative
corrections, which actually disappear from the subsequently derived
wave equations):
\subequations
\begin{eqnarray}
\label{e11a}
(i\partial_{x10}-m_1)G(x_1,x_2,x_1',x_2') &=& i\delta^4(x_1-x_1')
G_{20}(x_2-x_2') - \big [ \chi_{\xi}'(x_1-x_2)-\chi_{\xi}'(x_1-x_2')\big ]
\nonumber \\
& &\ \ \ \ \ \ \ \ \times G(x_1,x_2,x_1',x_2')\ ,
\\
\label{e11b}
(i\partial_{x20}-m_2)G(x_1,x_2,x_1',x_2') &=& i\delta^4(x_2-x_2')
G_{10}(x_1-x_1') - \big [ \chi_{\xi}'(x_2-x_1)-\chi_{\xi}'(x_2-x_1')\big ]
\nonumber \\
& &\ \ \ \ \ \ \ \ \times G(x_1,x_2,x_1',x_2')\ ,
\end{eqnarray}
\endsubequations
where the prime on $\chi$ designates the derivative with respect to
the temporal argument:
\begin{equation} \label{e12}
\chi_{\xi}'(x)\ \equiv\ \frac{\partial}{\partial x^0}\chi_{\xi}(x)\ .
\end{equation}
[$\chi_{\xi}'(x)$ is an odd function of $x^0$.]
\par
Continuing the procedure, one also obtains a second order differential
equation for $G$:
\begin{eqnarray} \label{e13}
& &(i\partial_{x10}-m_1)(i\partial_{x20}-m_2)G(x_1,x_2,x_1',x_2') \ =\
i^2\delta^4(x_1-x_1')\delta^4(x_2-x_2') + \big \{ i\chi_{\xi}''(x_1-x_2)
\nonumber \\
& & \ + \big[ \big(\chi_{\xi}'(x_1-x_2)-\chi_{\xi}'(x_1-x_2')\big)
\big(\chi_{\xi}'(x_2-x_1)-\chi_{\xi}'(x_2-x_1') \big)\big ]\big \}
G(x_1,x_2,x_1',x_2')\ .
\end{eqnarray}
\par
Wave equations satisfied by the Bethe$-$Salpeter wave function are
obtained by formally taking in the previous equations for $G$ the limit
$t\rightarrow \infty$ [Eq. (\ref{e6})], using the cluster
decomposition (\ref{e7}) and eliminating by integration on $X^{\prime 0}$
all the amplitudes $\phi_n$ but one \cite{gml}. One finds:
\subequations
\begin{eqnarray}
\label{e14a}
(i\partial_{10}-m_1)\phi(x_1,x_2) &=& -\big[\chi_{\xi}'(x)
+\frac{\lambda \pi}{2r}\big]\ \phi(x_1,x_2)\ ,\\
\label{e14b}
(i\partial_{20}-m_2)\phi(x_1,x_2) &=& -\big[-\chi_{\xi}'(x)
+\frac{\lambda \pi}{2r}\big]\ \phi(x_1,x_2)\ ,
\end{eqnarray}
\endsubequations
\begin{eqnarray} \label{e15}
(i\partial_{10}-m_1)(i\partial_{20}-m_2)\phi(x_1,x_2) &=& \big[i\chi_{\xi}''
(x) - \big(\chi_{\xi}^{\prime 2}(x)-(\frac{\lambda \pi}{2r})^2
\big) \big]\ \phi(x_1,x_2)\ .\nonumber \\
& &
\end{eqnarray}
In the Feynman gauge these equations reduce to those obtained by
Mugibayashi \cite{m} with the hamiltonian formalism.
\par
Equations (\ref{e14a})-(\ref{e14b}) yield for $\phi$ a single bound
state with the energy (\ref{e8p}) and wave function
\begin{equation} \label{e16}
\phi\ =\ Ce^{-i(m_1-m_2)x^0/2}e^{-i(m_1+m_2-\alpha/r)X^0}
\exp\big(i\chi_{\xi}(x)\big)\ ,
\end{equation}
with $C$, a constant.
\par
Equation (\ref{e15}) is best analyzed in euclidean space for the relative
time variable. Using Wick rotation \cite{n1,w} and then setting
$p_0=ip_0^E$ or $x^0=-i\tau$, Eq. (\ref{e15}) becomes (for simplicity
we consider the equal mass case $m_1=m_2=m$):
\begin{eqnarray} \label{e17}
-(\frac{P_0}{2}-m)^2\phi &=& - \frac{d^2}{d\tau^2}\phi
-\frac{\lambda}{2(\tau^2+r^2)^2}\ \big[(2+\xi)\tau^2+(2-\xi)r^2\big] \phi
\nonumber \\
& &\ \ \ \ \ \ \ \ \ \ -\big\{\ \big(\frac{\lambda \pi}{2r}\big)^2
-\lambda^2\big[ \frac{1}{r}\arctan(\frac{\tau}{r})-\frac{\xi \tau}
{2(\tau^2+r^2)}\big]^2\ \big\}\phi\ .
\end{eqnarray}
For large $|\tau|$ the potential
behaves as $-(1+\xi/2)\lambda^2 \pi/(2r|\tau|)$,
indicating that for $\xi>-2$ there are an infinite number of
normalizable solutions for any value of $\lambda$.
These additional solutions were identified with the abnormal solutions
of the Bethe$-$Salpeter equation. On the other hand, had we kept in the
above equation only the linear term in $\lambda$
(coming from the one-photon exchange contribution), we would have faced
the same situation as in the Wick$-$Cutkosky model \cite{w,c} or its
static analogue of Ref. \cite{otu}. In this case the additional
solutions exist only for $\lambda>1/(4(1+\xi/2))$.
\par
It is clear that these abnormal solutions do not appear as poles of the
Green's function because they do not satisfy the first order equations
(\ref{e14a})-(\ref{e14b}). At this stage the main objection one could
raise is the fact that Eqs. (\ref{e11a})-(\ref{e11b}) are not
fundamental equations of Quantum Field Theory and are specific to
the static model; they would not survive in more general cases
and hence only Eq. (\ref{e15}) would remain as a bound state equation,
leading to the
existence of abnormal solutions. Leaving aside for the moment the feature
that the
very existence and the energies of the abnormal solutions are manifestly
gauge dependent and in any event the latter could not appear as poles of
Green's functions \cite{ffh}, Eq. (\ref{e15}) suffers from the
following main drawback: in spite of appearances, it is {\it not} the
Bethe$-$Salpeter equation. The reason for
this is that the potential in Eq. (\ref{e15}) does not represent the
Bethe$-$Salpeter kernel. If this was the case, then we would conclude
that the series of multiphoton exchanges in the kernel stops at two
photon exchanges with a corresponding local expression in $x$-space. The
contribution of the two-photon exchange diagram to the kernel can be
explicitly calculated. In the Feynman gauge it reads:
\begin{eqnarray} \label{e18}
K_2(x_1,x_2,x_1',x_2') &=& \lambda^2 \theta(x_1^0-x_1^{\prime 0})
\theta(x_2-x_2^{\prime 0})\ \frac{e^{-im_1(x_1^0-x_1^{\prime 0})
-im_2(x_2^0-x_2^{\prime 0})}}{[(x_1-x_2')^2-i\epsilon]
[(x_2-x_1')^2-i\epsilon]} \nonumber \\
& & \ \ \ \ \ \ \ \ \ \ \ \ \times\ \delta^3({\bf x}_1-{\bf x}_1')\ \delta^3
({\bf x}_2-{\bf x}_2')\ .
\end{eqnarray}
This is not a local operator in the temporal variables and does not
cope with the $O(\lambda^2)$ terms of the potential of Eq. (\ref{e15}).
\par
More generally, the $O(\lambda^2)$ terms of Eq. (\ref{e13}), which are
multiplicative functions and act on the four arguments of the Green's
function, do not represent the explicit expression of the
Bethe$-$Salpeter kernel, which should act on two arguments only.
The latter can be calculated either from the Feynman diagrams or
iteratively by identifying its global action with the multiplicative
functions of Eq. (\ref{e13}). The general expression of the kernel
cannot, however, be put in a compact form and will not be reported here.
Its main feature is that the series of multiphoton exchange diagrams
does not stop at a finite order. Because the Bethe$-$Salpeter kernel
is directly related to the inverse of the Green's function [$K=G_0^{-1}
-G^{-1}$], the bound state spectrum of admissible states of the
Bethe$-$Salpeter equation should be the same as that of the Green's
function itself.
\par
The absence of abnormal states in the Bethe$-$Salpeter equation can
also be verified using gauge invariance of the lagrangian density
(\ref{e1}) (up to the gauge-fixing term) and working in the Coulomb
gauge. In this gauge, the expression
of the Green's function is obtained from Eqs. (\ref{e5})-(\ref{e5p}) by
the replacement of the photon propagator contribution by $-1/{\bf k}^2$.
This amounts to replacing $h_{\xi}$ by zero and $\chi_{\xi}$ by $\chi_C$
with:
\begin{equation} \label{e25}
\chi_C(x)\ =\ -\frac{\alpha}{2r}|x^0|\ .
\end{equation}
\par
For large time separations, the Green's function behaves as:
\begin{equation} \label{e25p}
{G_C}_{\stackrel{{\displaystyle \sim}}{t\rightarrow \infty}}
e^{-i(m_1+m_2-\alpha/r)t}\ ;
\end{equation}
it displays one bound state with energy given by Eq. (\ref{e8p}).
\par
The equivalent equations to Eqs. (\ref{e14a}), (\ref{e14b})
and (\ref{e15}) become:
\subequations
\begin{eqnarray}
\label{e26a}
(i\partial_{10}-m_1)\phi_C(x_1,x_2) &=& -\frac{\alpha}{r}\theta(-x^0)
\phi_C(x_1,x_2)\ ,\\
\label{e26b}
(i\partial_{20}-m_2)\phi_C(x_1,x_2) &=& -\frac{\alpha}{r}\theta(x^0)
\phi_C(x_1,x_2)\ ,
\end{eqnarray}
\endsubequations
\begin{equation} \label{e27}
(i\partial_{10}-m_1)(i\partial_{20}-m_2)\phi_C(x_1,x_2)\ =\ i\frac
{\alpha}{r}\delta(x^0)\phi_C(x_1,x_2)\ .
\end{equation}
\par
The latter equation is actually the Bethe$-$Salpeter equation in the
Coulomb gauge, because the instantaneity of the interaction in this gauge
reduces the Bethe$-$Salpeter kernel to the one-photon exchange diagram.
This equation has clearly one bound state solution, independently of
Eqs. (\ref{e26a}) and (\ref{e26b}), which confirms the fact the
Bethe$-$Salpeter equation should by itself contain the full information
about the bound state spectrum.
\par
The gauge transformation operator between two covariant gauges, or a
covariant gauge and
the Coulomb gauge, for the Green's function is obtained
by taking the ratio (in $x$-space) of their corresponding expressions;
it reduces to the ratio of the exponential factors with the $\chi$ and $h$
functions. Taking the large time separation one also deduces the gauge
transformation between the corresponding Bethe$-$Salpeter wave functions.
One finds:
\begin{equation} \label{e28}
\phi_{\xi'}(x_1,x_2)=e^{i[\chi_{\xi'}(x)-\chi_{\xi}(x)]} \phi_{\xi}
(x_1,x_2)\ ,\ \ \ \ \
\phi_C(x_1,x_2) = e^{i[\chi_C(x)-\chi_{\xi}(x)]} \phi_{\xi}(x_1,x_2)\ .
\end{equation}
\par
The above transformation laws, which reflect general properties of the
Bethe-Salpeter wave function \cite{bcm},
imply that if a bound state exists in one
gauge, it should also exist in any other gauge. This results from the
fact that the exponential factors involve the difference of two $\chi$'s
and hence do not qualitatively modify (in euclidean space) the dominant
asymptotic behavior of the wave functions.
\par
It can be checked that these transformations applied on Eqs.
(\ref{e14a})-(\ref{e14b}) yield the corresponding equations in the
new gauge. However, the equivalence of the Bethe$-$Salpeter equations of
two different gauges cannot be established in strong
form. A general feature of the Bethe$-$Salpeter equation is that it is only
weakly invariant under gauge transformations, that is that its
(infinitesimal) variation is proportional to the starting wave equation.
[This behavior can be established by writing the equation in the
form $G^{-1}\phi=0$ and then using the infinitesimal gauge transformations
of $G$ and $\phi$.] In particular, using in Eq. (\ref{e27})
transformation (\ref{e28}) one obtains an equation that is weakly
equivalent to the Bethe-Salpeter equation of the covariant gauge.
Nevertheless, the general property of the invariance of the bound state
spectra under gauge transformations \cite{ffh}, explicitly verified
on transformations (\ref{e28}), ensures us that the
latter equation has also one bound state.
\par
The previous calculations can also be repeated with a massive photon with
mass $\mu$.
For large time separations, the Green's function behaves as:
\begin{equation}  \label{e29}
G_{\ \buildrel {\displaystyle \sim }\over {t \to \infty}}\ e^{
 -i(m_1+m_2-\alpha e^{-\mu r}/r )t  }\ .
\end{equation}
We deduce that there is only one bound state, the normal one.
\par
In summary, the exact expression of the four-point Green's function
in the static model does not display any abnormal state. In the
one-photon exchange approximation (in covariant gauges), the situation
is similar to that of the Wick$-$Cutkosky model, where abnormal solutions
appear for sufficiently large values of the coupling constant. The present
study shows that the multiphoton exchange contributions sum up to
sweep away the previous phenomenon and to reestablish a normal structure
of the bound state spectrum. Since the phenomenon of the abnormal
states is solely due to the relative time excitations, the results
obtained in the static model, where spacelike motion is frozen,
should survive in the more general case of four-dimensional theory.
\par

\end{document}